\documentclass[twocolumn,showpacs,aps,prl]{revtex4}
\usepackage{amsmath,amsfonts,bm,graphicx}

\newcommand{\G}{\Gamma}
\newcommand{\g}{\gamma}
\newcommand{\eps}{\epsilon}
\newcommand{\w}{\omega}
\newcommand{\s}{\sigma}
\renewcommand{\t}{\tau}
\newcommand{\ixe}{\rangle\!\langle}
\newcommand{\phd}{\phantom{\dagger}}
\newcommand{\Pj}{\mathcal{P}}

\begin{document}

\title{Dynamical symmetry breaking in transport through molecules}

\author{Andrea Donarini}

\affiliation{Theoretische Physik, Universit\"{a}t Regensburg,
93040 Regensburg, Germany}

\author{Milena Grifoni}

\affiliation{Theoretische Physik, Universit\"{a}t Regensburg,
93040 Regensburg, Germany}

\author{Klaus Richter}

\affiliation{Theoretische Physik, Universit\"{a}t Regensburg,
93040 Regensburg, Germany}

\date{\today}

\begin{abstract}
We analyze the interplay between vibrational and electronic
degrees of freedom in charge transport across a molecular
single-electron transistor. We focus  on the wide class of
molecules which possess quasi-degenerate vibrational eigenstates,
while no degeneracy occurs for their anionic configuration. We
show that the combined effect of a thermal environment and
coupling to leads, involving tunneling events charging and
discharging the molecule, leads to a dynamical symmetry breaking
where quasi-degenerate eigenstates acquire different occupations.
This imbalance gives rise to a characteristic asymmetry of the
current versus an applied gate voltage.

\end{abstract}

\pacs{85.65.+h, 85.85.+j, 73.63.–b}

\maketitle

Molecular electronics is a promising answer to the demand of
miniaturization, reproducibility and flexibility of electronic
devices. Starting with the pioneering work of Aviram and Ratner
\cite{avi-cpl-74}, and especially with the first single-molecule
measurement by Reed {\it et al.\ } \cite{ree-sci-97}, molecular
electronics has become an active research field both
experimentally and theoretically \cite{reviews}. Still,
fundamental questions on the peculiar nature of single-molecule
junctions and their novel functionalities remain open. Among
others they involve the unique electromechanical properties  of
molecular junctions which render these distinctly different from,
e.g., semiconductor quantum dots.

Recent research on this particular topic has revealed a number of
interesting effects such as shuttling instabilities
\cite{gor-prl-98,nov-prl-03,par-nat-00}, Franck-Condon blockade
\cite{koc-prl-05,nowack-05} and,  more generally, conformational
\cite{sam-prb-96,don-sci-01,ran-prb-04} or vibronic~
\cite{bra-prb-03,tro-jcp-03,mit-prb-04,ciz-cjp-05} signatures in
the electron transport characteristics. In this respect, however,
the role of coherences in transport due to quasi-degenerate levels
has not been highlighted.

\begin{figure}
  \includegraphics[width=55mm,angle=-90]{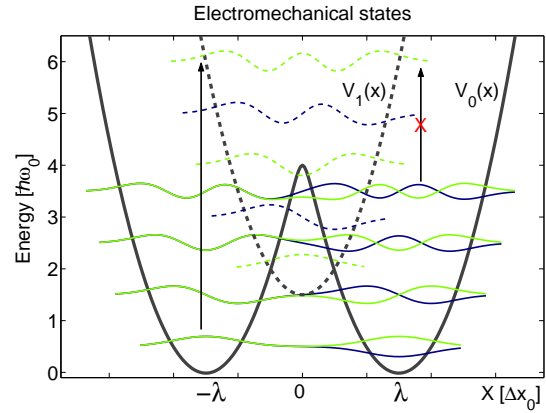}
  \caption{(color online) Electromechanical states of the molecule.
  Thick black lines indicate the adiabatic potentials $V_0$ and $V_1$ for the neutral (solid)
  and singly-charged (dashed) molecule. Thin green (blue) lines denote the even (odd)
  vibrational eigenfunctions (for the two potentials) displayed
  with a vertical shift equal to their corresponding energy eigenvalue.
  Energies are given in units of $\hbar \omega_0$ and lengths in terms of the
  zero-point motion $\Delta x_0 = \sqrt{\hbar/m\omega_0}$.
  The arrows mark examples of allowed and forbidden transitions.}
  \label{fig1}
\end{figure}

In this Letter we consider charge transport through a molecular
junction weakly contacted to leads.  At low enough temperatures
charging effects prevent that more than one excess electron at a
time can populate the molecule. The weak coupling ensures that the
potential drop between source and drain leads is concentrated in
the contact region, and no substantial electric field acts
directly on the molecule. We focus on the ubiquitous, but so far
poorly investigated, case of adiabatic vibronic potentials with
more than one stable configuration as {\it e.g.} found in
oligoparaphenylenes. In particular, we address molecules
characterized by eigenstate dubletts in the neutral configuration
and non-degenerate anionic states, or vice versa. This is sketched
in Fig.\ \ref{fig1}. which can represent, e.g., the adiabatic
potential energy surfaces of biphenyl-based molecules as a
function of the dihedral angle (here playing the role of the $x$
coordinate) between the phenyl rings \cite{ciz-cjp-05}. These
molecules have been studied in recent transport experiments
\cite{dad-nat-05,ven-nat-06}. Specifically, in \cite{ven-nat-06}
for the first time quantitatively, the connection between the
angular conformation and the electrical conductance has been
proved for several biphenyl-based molecules.

We show that dynamical symmetry breaking (DSB), where
quasi-degenerate eigenstates are differently occupied, may occur
and affect transport, while in the absence of couplings to the
leads these states are equally populated at finite temperature,
and hence the system does not prefer a definite parity.
Temperature, on the other hand, should  not exceed a critical
value above which DSB is lost due to dephasing, caused by a
thermal bath. Solving the master equation for the reduced density
matrix including coherences, we demonstrate the possibility of
detecting DSB in the current through the molecule under different
bias and gating conditions.

To address transport in the single-electron regime, where only the
electronic ground states $|0\rangle$ of the neutral and
$|1\rangle$ of the anionic configuration are involved, we describe
the molecule through the Hamiltonian
\begin{equation}
H_{\rm S} =
 \frac{p^2}{2m} + |0\ixe
 0|\,V_0(x)+|1\ixe1|\,V_1(x)
\end{equation}
in terms of the corresponding adiabatic potential energy surfaces
$V_0(x)$, $V_1(x)$ associated with the softest mode coordinate
$x$. The adiabatic potentials and the related vibrational
eigenfunctions are sketched in Fig.\ \ref{fig1}. The neutral
molecule is characterized by the  double well potential $V_0$ with
minima located at $\pm\lambda$. For large enough barriers, i.e.\
exponentially small tunnel splitting,  the low-lying eigenstates
are organized in pairs of (quasi-)degenerate wave functions. The
anion is modelled by a single harmonic potential $V_1$ with
frequency $\omega_1$, centered at the origin of the coordinate
system (dashed line in Fig.~1).

We express the system coupled to leads and in contact with a thermal bath
through the Hamiltonian
\begin{equation}
    H =
    H_{\rm S} + H_{\rm L}+H_{\rm R}
    + H_{\rm B} + V_{\rm T} + V_{\rm SB}\;.
    \label{eq:Htot}
\end{equation}
The leads are described as  reservoirs
of non-interacting quasi-particles in terms of $H_{\rm
L} + H_{\rm R} =\sum_{k;\alpha={\rm L,R}}
    (\eps_{k\alpha}-\mu_{\alpha})c_{k\alpha}^{\dagger}c_{k\alpha}^{\phd}$,
where $\mu_{\rm L,R} = \pm \Delta V/2$  accounts for the effects
of a symmetrically applied bias voltage and
$c_{k\alpha}^{\dagger}$ ($c_{k\alpha}$) creates (destroys) an
electron in the lead $\alpha$. Transfer of electrons is mediated
by the tunneling Hamiltonian $ V_{\rm T} = v \sum_{k;\alpha={\rm
L,R}}
 (|0\ixe 1|c_{k\alpha}^{\dagger}+ |1\ixe
 0|c_{k\alpha}^{\phd})\,.
$
Finally,  $H_B$ describes a thermal bath of harmonic
oscillators coupled linearly to the displacement coordinate $x$ via the
system-bath Hamiltonian $V_{\rm SB}=\hbar \tilde g  x \sum_q
(d_q^\dagger + d_{q})$, where $d_{q}^{\dagger}$ and $d_q$ are bosonic
creation and destruction operators associated with the oscillator with
energy $\hbar \omega_q$.

Since we are interested only in the low-energy states, we
approximate $V_0(x)$ by
two harmonic wells centered at $\pm \lambda$ and with
frequency $\omega_0$. Formally, this is achieved by
projecting the Hamiltonian $H$ on a smaller Hilbert space in which we
can define the identity operator as
\begin{equation}
\openone = |1\ixe 1| + |0\ixe 0|(\Pj_+ + \Pj_-)\;.
\label{eq:identity}
\end{equation}
\begin{figure*}
\includegraphics[width=0.95\textwidth, clip=on]{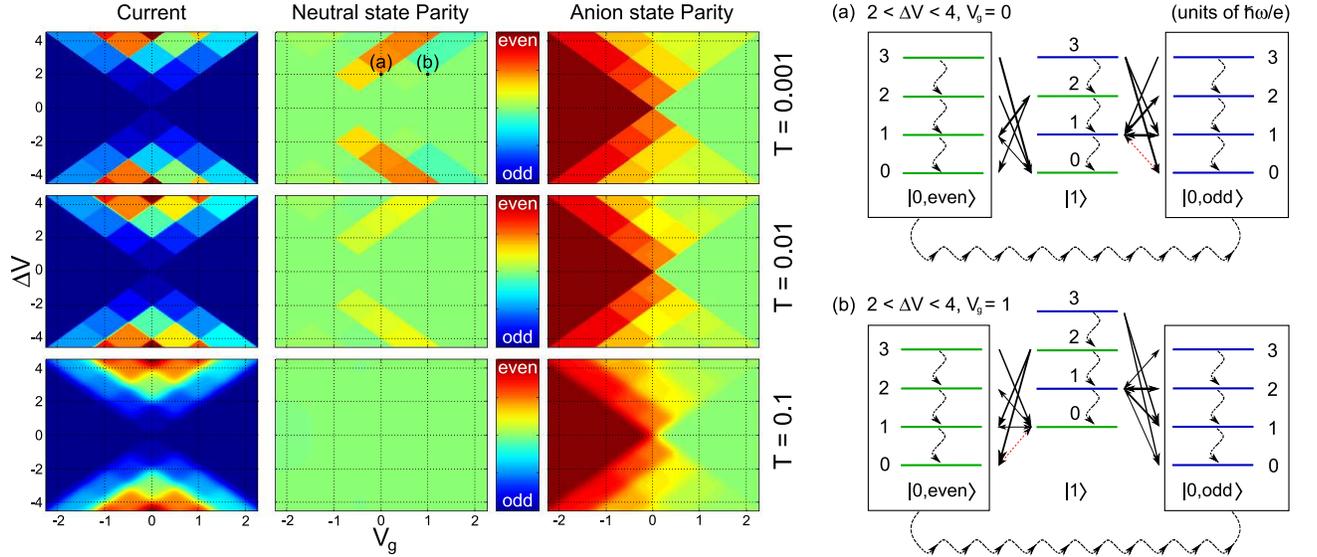}
\caption{(color online)
 Left hand side. Onset of the dynamical symmetry
    breaking with decreasing temperature (in units of
    $\hbar\omega_0/k_{\rm B}$) from bottom to top.
    The columns show the {\it absolute} value of the stationary current and the
    parity of the neutral and anionic state as a function of the gate voltage
    $V_g$ and the bias $\Delta V$ across the molecular
    junction, both in units of $\hbar\omega_0 /e$.
    The other parameters are: $\gamma = 0.01\,\omega_0$,
    $\Gamma_{L,R}=0.1\,\omega_0$, $\lambda=3\,\sqrt{\hbar/m\omega_0}$.
 Right hand side. Rates scheme at the even (a) and odd (b)
    transition points (corresponding to the
    points marked by (a) and (b) in the upper middle panel of
    the figure to the left. The horizontal lines mark the electromechanical states of the neutral
    and anionic configuration with green (blue) color representing the even (odd) parity.
    Straight arrows represent transitions due to the coupling to
    the leads: the rate is qualitatively expressed by the thickness of
    the line. Dashed arrows show transitions lifting the blocking character of the ground states
    and are crucial for the understanding of the symmetry breaking transition.
    Wavy arrows represent the effect of the phononic bath.
  }
  \label{fig2}
\end{figure*}

Here  $\Pj_{\pm} = \sum_{n=0}^{N}|n,\pm\ixe n,\pm|$ are projection
operators on the states right/left to the barrier. They are
defined in terms of the shifted harmonic oscillator eigenvectors
$|n,\pm\rangle \equiv \exp(\mp i \lambda p/ \hbar)|n\rangle$. The
condition of negligible overlap between these eigenfunctions reads
$\Pj_+\,\Pj_- = \Pj_-\,\Pj_+ = 0$. In the Hilbert space defined by
identity \eqref{eq:identity} we obtain as effective system
Hamiltonian
\begin{equation}
\begin{split}
 H_{\rm S}^{\rm eff} =& |0 \ixe 0| \sum_{\t= \pm }\left[
 \Pj_{\t}
 (\tfrac{1}{2}+d_{\t}^{\dagger}d_{\t}^{\phd})
 \Pj_{\t}\right]\hbar\w_0 + \\
 & \qquad +|1 \ixe 1|
 \left[(\tfrac{1}{2}+d^{\dagger}d)\hbar\w_1+eV_g-\eps_a\right]\;,
 \end{split}
 \end{equation}
where $d^\dagger_{\pm}$ and $d^\dagger$ are operators for the
neutral and charged configurations, $V_g$ is the gate voltage, and
$\eps_a$ the electron affinity. Along similar lines we find
for the effective tunneling and system-bath Hamiltonians
 \begin{equation}
 \begin{split}
 V_{\rm T}^{\rm eff} =& \; v \sum_{k;\alpha={\rm L,R}}
 \left[ |0\ixe 1|c^{\dagger}_{k\alpha}(\Pj_+ + \Pj_-) +
 \,h.c.\right]\;,\\
 V_{\rm SB}^{\rm eff} =& \;
  \hbar g \sum_q (d_q^{\phd} + d_q^{\dagger})
  \Big\{
   \delta |1 \ixe 1|(d + d^{\dagger})+\\
   & \quad +|0 \ixe 0|\sum_{\t = \pm}[\Pj_{\t}(d_{\t}^{\phd}+d_{\t}^{\dagger}+\t 2 \lambda
   )\Pj_{\t}]
  \Big\}\;,
\end{split}
\end{equation}
with $g=\tilde{g}\sqrt{2\hbar/(m\w_0)}$ and $\delta =
\sqrt{\w_1/\w_0}$. For transport, we seek
stationary solutions of the Liouville equation
\begin{equation}
\dot \sigma (t)=-\frac{i}{\hbar} {\rm Tr}_{\rm leads + bath}\{[H^{\rm eff},W(t)]\}
\label{eq:Liouville-1}
\end{equation}
for the reduced density matrix (RDM) $\sigma (t) := {\rm Tr}_{\rm
leads + bath}\{W(t)\}$, where $W(t)$ is the total density matrix
associated with the effective Hamiltonian corresponding to that in Eq.~(\ref{eq:Htot}). Treating
the interactions $V_{\rm T}^{\rm eff}$ and $V_{\rm SB}^{\rm eff}$
as perturbations, we rewrite Eq.~(\ref{eq:Liouville-1})
 in terms of  three contributions: $\dot{\s}(t)  = \mathcal{L} \sigma (t) =
  (\mathcal{L}_{\rm coh}+\mathcal{L}_{\rm tun}+\mathcal{L}_{\rm damp})
  \,\s(t)$. The coherent part of the differential equation has the usual form
$ \mathcal{L}_{\rm coh}\,\s(t) = -(i/\hbar)[H_{\rm S}^{\rm eff},\s(t)] $
describing the evolution of the isolated molecule. The
coupling to the leads/bath gives rise to the driving/damping
terms ${\cal L}_{\rm tun}\sigma(t)$ and ${\cal L}_{\rm
damp}\sigma(t)$.

 We now perform the following standard
approximations: we treat the leads as thermal reservoirs at
equilibrium temperature $T$, consider the coupling to the leads
up to second order in $V_{\rm T}^{\rm eff}$ and, being
interested in the stationary solution, neglect non-Markovian
contributions to the equation of motion. Furthermore, neglecting off-diagonal
elements of the RDM between anionic and neutral configurations, as
well as between non-degenerate eigenstates, and disregarding
non-energy conserving contributions, the anionic and neutral state
components read
 \begin{eqnarray}
 \left(\mathcal{L}_{\rm tun}\,\s\right)_{11}
 &=& \! \! \sum_{\alpha ={\rm L,R};\t =\pm }\Big[
  2(\G^{\alpha}_{\rm in}\,\s_{00}\,\Pj_{\t}+
  \Pj_{\t}\,\s_{00}\,{\G^{\alpha}_{\rm
  in}}^{\dagger}) - \nonumber\\
 &  & \qquad\qquad - (\Pj_{\t}\,\G^{\alpha}_{\rm out}\,\s_{11}+
  \s_{11}\,{\G^{\alpha}_{\rm out}}^{\dagger}\Pj_{\t})
  \Big]\;,\nonumber \\
\left(\mathcal{L}_{\rm tun}\,\s\right)_{00}^{\t\t'} &=&
\sum_{\alpha
 {\rm =L,R}}\Pj_{\t}\Big[
  \G^{\alpha}_{\rm out}\,\s_{11}+
  \s_{11}\,{\G^{\alpha}_{\rm out}}^{\dagger} - \\
&   &  \qquad\qquad -2(\G^{\alpha}_{\rm in}\,\s_{00}+
  \s_{00}\,{\G^{\alpha}_{\rm
  in}}^{\dagger})\Big]\,\Pj_{\t'} \nonumber \;,
\label{eq:rho-tun}
 \end{eqnarray}
with rate matrices for tunneling into/out of the molecule
\begin{eqnarray}
 \G^{\alpha}_{\rm in} &= & \frac{\G_{\alpha}}{2}\sum_{m,n,\t =\pm}|m\rangle
\langle n,\t| f(eV_g
 +\epsilon_{nm})\langle m|n,\t
 \rangle
 \;,\nonumber \\
 \G^{\alpha}_{\rm out} &=&
 \frac{\G_{\alpha}}{2}\sum_{m,n,\t=\pm}|n,\t\ixe m| -
 {\G^{\alpha}_{\rm in}}^{\dagger} \;.
\label{eq:rates}
\end{eqnarray}
Here $f(\epsilon)$ is the Fermi function, and $\G_{\alpha} =
(2\pi/ \hbar) D_{\alpha} |v|^2$ is the bare electronic rate, with
$D_\alpha$ the density of states in lead $\alpha$ at the Fermi
energy. For convenience we assumed equal frequencies $\w_0 = \w_1$
of the neutral and anionic potential, and hence
$\epsilon_{nm}=\hbar\omega_0(m-n)$; calculations for $\w_0 \neq
\w_1 $ yielded no qualitative changes. Furthermore we shifted the
electrochemical potential $eV_g - \epsilon_a \rightarrow eV_g $ in
the argument of the Fermi function. The overlap matrix elements
$\langle m|n,\t \rangle$ of the vibrational states of the neutral
and charged configurations, known as Franck-Condon factors,
determine, together with energy resonance conditions, the
transition rates (\ref{eq:rates}) and thus the transport
characteristics of the junction. While the Franck-Condon factors
are fixed by the adiabatic potentials, the bias and gate voltages
influence the resonance conditions.

To describe relaxation and dephasing, we consider the case of an
Ohmic bath with spectral density $J(\omega)=m\gamma\omega$, where
$\gamma$ is the damping coefficient  \cite{wei-book-99}.  Along
the same lines as for the tunneling term we find
 \begin{eqnarray}
 \left(\mathcal{L}_{\rm damp}\,\s\right)_{11} &=&
  -\frac{i\g}{2\hbar}[x,\{p,\s_{11}\}] - \nonumber \\
 & & - \frac{\g m \omega_0}{\hbar}(\bar{N}+\tfrac{1}{2})[x,[x,\s_{11}]]\nonumber \;,\\
\left(\mathcal{L}_{\rm damp}\,\s\right)_{00}^{\t\t'}
&=&-\frac{i\g}{2\hbar}\Pj_{\t}[x,\{p,\s_{00}^{\t\t'}\}]\Pj_{\t'} \label{eq:rho-diss} - \\
 &&  - \frac{\g m
 \omega_0}{\hbar}(\bar{N}+\tfrac{1}{2})\Pj_{\t}[x,[x,\s_{00}^{\t\t'}]]\Pj_{\t'}- \nonumber \\
 && - 8\g\frac{k_{\rm B}T}{\hbar\omega_0}\lambda^2(\Pj_+\,\s_{00}^{\t\t'}\,\Pj_- +
 \Pj_-\,\s_{00}^{\t\t'}\,\Pj_+)\;, \nonumber
 \end{eqnarray}
where ${\bar N} (\omega)$ is the Bose distribution function. Note
the occurence of a term responsible for pure inter-well  dephasing
and proportional to $T$ and $\lambda^2$ in the expression for
$\left(\mathcal{L}_{\rm damp}\,\s\right)_{00}$.

Since we are interested in the long-time properties, we look for
solutions of the stationary problem $\mathcal{L}\,\s^{\rm
stat}\!=\!  0$. Given the stationary RDM $\s^{\rm stat}$ we
calculate the  stationary current as the trace over the system
degrees of freedom of the left/right current operators, $  I^{\rm
stat}\! =\! {\rm Tr}_{\rm mech} [\s^{\rm stat}\hat{I}_{\rm L}]
        \!  =\!  {\rm Tr}_{\rm mech} [\s^{\rm stat}\hat{I}_{\rm R}]\;,$
with, e.g.,
\begin{equation}
\begin{split}
  \hat{I}_{\rm L} = \sum_{\t =\pm}\Big[&
  2|0\ixe 0|(\Pj_{\t}\,\G_{\rm in}^{\rm L}+
  {\G_{\rm in}^{\rm L}}^{\dagger}\,\Pj_{\t})- \\
  &-|1\ixe 1|(\Pj_{\t}\,\G_{\rm out}^{\rm L}
  + {\G_{\rm out}^{\rm L}}^{\dagger}\,\Pj_\t)\Big]\;.
\end{split}
\end{equation}
We stress that not only populations but also coherences of the RDM
in the $(\pm)$ basis contribute to $I^{\rm stat}$.

In the first column of the left side of Fig.\ \ref{fig2} we
present the results of our calculation of the gate- and
bias-dependent current for decreasing temperature (from bottom to
top). Besides the evidence of a Franck-Condon blockade
\cite{koc-prl-05,nowack-05} (where due to an exponential
suppression of the Franck-Condon factors $\langle m|n,\t \rangle$
transport is blocked and the Coulomb diamonds no longer close) and
current steps due to phononic excitations, Fig.\ \ref{fig2} shows
an increasing asymmetry in the gate voltage $V_g$ with decreasing
temperature. To  understand the origin of this asymmetry, we
depict in the second and third column of the left side of Fig.\ 2
the parities of the neutral and anionic state defined as
\begin{eqnarray}
 P_0 &=& \sum_{n=0}^{N}\Big[\langle 0,e,n |\s| 0,e,n \rangle - \langle 0,o,n |\s| 0,o,n
 \rangle\Big] \;,\\
 P_1 &=& \sum_{n=0}^{\infty}\Big[\langle 1,2n |\s| 1,2n \rangle - \langle 1,2n+1 |\s| 1,2n+1
 \rangle\Big]\;,\nonumber
\end{eqnarray}
where the even/odd states of the neutral molecule are
\begin{equation}
 |0,e/o,n \rangle = \frac{1}{\sqrt{2}}(|n,+ \rangle \pm (-1)^n |n,- \rangle)\;.
 \label{eq:basis}
\end{equation}
In the parameter region where the current becomes asymmetric with
respect to $V_g$, degenerate states are {\em differently}
populated, i.e., a dynamical symmetry breaking (DSB) in the
occupation of the even/odd states occurs.
 In the regions of defined parity (warm colors for
even, cold for odd) the system exhibits spatial coherences in the
$(\pm)$ basis, cf.\ Eqs.~(\ref{eq:rho-tun}),(\ref{eq:rho-diss}).
In contrast, we find that in the even/odd basis coherences are
decoupled from populations and vanish for the stationary density
matrix. The onset of this even/odd parity regions in the gate-bias
voltage plane can be understood in terms of transition rates
between the different vibrational states of the molecule (see the
right side of Fig.\ \ref{fig2}).  At zero bias, energy
conservation prevents any transition \emph{from} the neutral (even
or odd) or anionic ground states to occur. These states are thus
blocking states. In particular for positive gate voltages the
molecule is neutral and dephasing ensures equilibration of the
even and odd populations of the degenerate ground state ($P_0=0$).
At negative gate voltages the molecule is charged and dissipation
ensures relaxation predominantly to the  anionic ground state
($P_1 \approx 1$).

The situation  changes for combinations of bias and gate voltages
that allow current to flow: the electromechanical system is
maintained out of equilibrium
by the applied voltage. Now the parity of the anionic state fluctuates due
to population of the higher excited states, and  the distribution
of the populations  no longer corresponds to the thermal
distribution in equilibrium with the bath.

The sign of the neutral-state parity  depends on the particular
rate configuration. Though a quantitative prediction relies on the
solution for $\s^{\rm stat}$, it is possible to understand the
sign of the neutral state parity with simple arguments. In the
right side of Fig.\ \ref{fig2} a  representation of the
electromechanical rates for two specific cases is reported.
Despite the complexity of the scheme, only  few lines are crucial:
these are the two red dashed arrows representing the rates which
lift the blocking character of the neutral ground states. In the
case (a) the neutral odd ground state can be depopulated resulting
in an even parity of the stationary state. In case (b) the neutral
even ground state is involved leading to an odd parity of the
stationary state. These unblocking rates represent the bottle-neck
in the DSB and are competing with the dephasing generated by the
heat bath (wavy line joining the even and odd sector of the
neutral state). The intensity of the unblocking rates depends on
the Franck-Condon factors of the vibrational wave functions
involved. A comparison with the dephasing rate allows to estimate
the transition temperature $T_{\rm tr}$ as $k_{\rm B}T_{\rm tr} =
\hbar \w_0 \G_{\rm ub}/ (4 \lambda^2 \g)$ , where $\G_{\rm ub}$ is
the unblocking rate (i.e.\ the \emph{depopulation} rate of the
relevant neutral ground state). The even transition (a) involves a
larger $\G_{\rm ub}$: it is more robust and occurs at higher $T$
(compare first and second panel in the central column of the left
side of Fig.~\ref{fig2}).

\begin{figure}[h!]

\end{figure}

To summarize,  we  analyzed the dynamics of a molecular junction
in the single-electron transport regime. The molecule possesses
quasi-degenerate vibrational eigenstates in the neutral
configuration and no degeneracy in the anionic case. Tunneling
processes charging and discharging the molecule preserve the
parity of the wavefunctions. As a consequence, unequal occupation
of degenerate molecular neutral states occurs. An explanation of
this effect in terms of unblocking rates is given.

We acknowledge financial support by the DFG within the research
school GRK 638. We thank in particular M.~\v{C}\'{\i}\v{z}ek for
inspiring discussions.

\end{document}